# Geometrization of spin systems using cycle expansions

Ronnie Mainieri

*Theoretical Division, MS B213, Los Alamos National Laboratory,*
*Los Alamos, NM 87545*

## Abstract

It is shown that a spin system with long range interactions can be converted into a chaotic dynamical system that is differentiable and low-dimensional. The thermodynamic limit of the spin system is then equivalent to studying the long term behavior of the dynamical system. Cycle expansions of chaotic systems (expansion of the Fredholm determinant) are then used to study the thermodynamic limit. By considering the smooth dynamical system, it is possible to converge to the thermodynamic limit faster than with transfer matrices.

Cycle expansions are an efficient method to compute the properties of ergodic systems [1]. They have been successfully applied to chaotic dynamical systems [1, 2], spin systems in statistical mechanics [3, 4], and to the semiclassical quantization of chaotic systems [5, 6]. In a cycle expansion, periodic orbits of the system are defined and each orbit is given a weight (such as the stability of an orbit, or the energy of a configuration). The weights are then combined into an expansion that uses shorter orbits to estimate the weight of longer orbits. The more orbits added to the expansion, the more accurate the estimates. The great advantage of cycle expansions is that analytic understanding of the system (on symmetries, or on partly integrable regions, for example) can be combined with numerical results into a rapidly converging expansion. In the case of dynamical systems what is expanded is the Fredholm determinant of the operator that determines the invariant measure of the system — the Perron-Frobenius operator. This operator is defined in such a way that the determinant has no singularities, which leads to expansions that converge faster than any exponential in the length of the longest periodic orbit. For statistical mechanics the transfer matrix is very similar to the Perron-Frobenius operator, but it leads to a determinant with singularities. And in quantum mechanics it is not known, in general, how to construct the operator equivalent to the Perron-Frobenius or transfer matrix (although Cvitanović and Vattay have solved the semiclassical version of this problem [7]).

I would like to show that it is possible to construct a Fredholm determinant for spin systems. I will do this by transforming the spin system into a low-dimensional dynamical system and then constructing its Fredholm determinant. Using the known results of chaotic systems, this means that expansions for thermodynamic quantities will converge faster (both qualitatively and quantitatively) than the exponentially fast convergence obtained using transfer matrices. There are many advantages of carrying out this procedure for a spin system: it allows the use of differentiable methods, before difficult due to the discrete nature of spins; and it gives a general framework in which to construct other Fredholm determinants, as there are few restrictions on the spin interaction.

The use of methods from statistical mechanics of spin systems in the study of dynamical systems — the thermodynamic formalism — is now well established. It was pioneered by Sinai [8], Bowen [9], and Ruelle [10] in the early seventies and it had its great success in physics after its application to the Feigenbaum period doubling route to chaos [11, 12]. In contrast, the methods of dynamical systems have not been used to study spin systems. With the geometrization of spin systems this reverse route becomes possible. As a first application, geometrization provides a counter-example to the belief that, aside from exactly solved models, the thermodynamic limit can only be approached exponentially fast.

The basic idea in the geometrization of spin systems is not to consider the shift of spin configurations as the dynamics, but instead to consider the dynamics of a few observables (such as the interaction energy or the magnetization). To explain how this is done, I will review the transfer operator (a generalization of the transfer matrix) which acts on observables (functions of configurations). Then the $\mathcal{L}$ operator will be introduced, similar to the transfer operator, but acting in a more restricted space, that of functions that depend only on the effective field at a point. This difference turns out to be crucial. To study the spectrum of the $\mathcal{L}$ operator one uses the techniques of cycle expansions. A difficulty in computing the trace of $\mathcal{L}^n$ is solved by considering a functional derivative involving the interaction. As an example of the advantages of using the $\mathcal{L}$ operator, I show that the thermodynamic limit is approached faster than any exponential.

To geometrize spin systems, the interactions are assumed to be translationally invariant and smooth. The spins $\sigma_k$ will only assume a finite number of values. For simplicity I will take the interaction $\phi$ among the spins to depend only on pairwise interactions,

$$\phi(\sigma) = \phi(\sigma_0, \sigma_1, \sigma_2, \ldots) = J_0 \sigma_0 + \sum_{n>0} \delta_{\sigma_0, \sigma_n} J_1(n) \tag{1}$$

and the possible values of a spin $\sigma_n$ to be in $\Omega_0 = \{+, -\}$. For the one-dimensional Ising model, $J_0$ is the external magnetic field and $J_1(n) = 1$ if $n = 1$ and $0$ otherwise. For an exponentially decaying interaction $J_1(n) = e^{-\alpha n}$.

The transfer operator $\mathcal{T}$ was introduce by Kramers and Wannier [13] to study the Ising model on a strip and concocted so that the trace of its n-th power is the partition function $Z_n$ of system when one of its dimensions is $n$. The method can be generalized to deal with any finite-range interaction. If the range of the interaction is L, then $\mathcal{T}$ is a matrix of size $2^L \times 2^L$. The longer the range, the larger the matrix. When the range of the interaction is infinite the matrix definition becomes cumbersome, and it is simpler to define the $\mathcal{T}$ operator by its action on an observable g. Just as the observables in quantum mechanics, g is a function that associates a number to every state (configuration of spins). The energy density and the average magnetization are examples of observables. From this equivalent definition one can recover the usual transfer matrix by making all quantities finite range. For a semi-infinite configuration $\sigma = \{\sigma_0, \sigma_1, \ldots\}$:

$$\mathcal{T} g(\sigma) = g(+ \vee \sigma) e^{-\beta \phi(+\vee\sigma)} + g(- \vee \sigma) e^{-\beta \phi(-\vee\sigma)} . \tag{2}$$

By $+ \vee \sigma$ I mean the configuration obtained by prepending $+$ to the beginning of $\sigma$ resulting in the configuration $\{+, \sigma_0, \sigma_1, \ldots\}$. When the range becomes infinite, the tr$\mathcal{T}^n$ is infinite and there is no longer a connection between the trace and the partition function for a system of size $n$ (this is a case where matrices give the wrong intuition). Ruelle [14] generalized the Perron-Frobenius theorem and showed that even in the case of infinite range interactions the largest eigenvalue of the $\mathcal{T}$ operator is related to the free-energy of the spin system and the corresponding eigenvector is related to the Gibbs state. By applying $\mathcal{T}$ to the constant observable $u$ that returns 1 for any configuration, the free energy f is computed as

$$-\beta f(\beta) = \lim_{n \to \infty} \frac{1}{n} \ln \|\mathcal{T}^n u\| \; . \quad (3)$$

To construct a smooth dynamical system that reproduces the properties of $\mathcal{T}$, one uses the phase space reconstruction technique of Packard *et al.* [15] and Takens [16], and introduces a vector of state observables $x(\sigma) = \{x_1(\sigma), \ldots, x_D(\sigma)\}$. To avoid complicated notation I will limit the discussion to the example $x(\sigma) = \{x_+(\sigma), x_-(\sigma)\}$, with $x_+(\sigma) = \phi(+ \vee \sigma)$ and $x_-(\sigma) = \phi(- \vee \sigma)$. We also restrict all observables to be functions only of the state observables $\{x_k(\sigma)\}$. That is, the only accepted observables are those g for which, for all configurations $\sigma$, there exist an analytic function G such that $G(x_1(\sigma), \ldots, x_D(\sigma)) = g(\sigma)$. This at first seems a severe restriction as it may exclude the eigenvector corresponding to the Gibbs state. It can be checked that this is not the case by using the formula given by Ruelle [17] for this eigenvector. A simple example where this formalism can be carried out is for the interaction $\phi(\sigma)$ with pairwise exponentially decaying potential $J_1(n) = a^n$ (with $|a| < 1$). In this case $\phi(\sigma) = \sum_{n>0} \delta_{\sigma_0, \sigma_n} a^n$ and the state observables are $x_+(\sigma) = \sum_{n>0} \delta_{+, \sigma_n} a^n$ and $x_-(\sigma) = \sum_{n>0} \delta_{-, \sigma_n} a^n$. These two observables always add up to the same constant. In this case the observable $x_+$ gives the energy of $+$ spin at the origin, and $x_-$ the energy of a $-$ spin.

As an example, assume that the phase space of the dynamical system is given by the two observables $\{x_+(\sigma), x_-(\sigma)\} = \{\phi(+ \vee \sigma), \phi(- \vee \sigma)\}$. In terms of these observables the transfer operator can be re-expressed

$$\mathcal{T} G(x(\sigma)) = \sum_{\eta = +, -} G\left(x_+(\eta \vee \sigma), x_-(\eta \vee \sigma)\right) e^{-\beta x_\eta(\sigma)} \; . \quad (4)$$

In this equation the only reference to the configuration $\sigma$ is when computing the new observable values $x_+(\eta \vee \sigma)$ and $x_-(\eta \vee \sigma)$. The iteration of the function that gives these values in terms of $x_+(\sigma)$ and $x_-(\sigma)$ is the dynamical system that will reproduce the properties of the spin system. For the simple exponentially decaying potential this is given by two maps, $F_+$ and $F_-$, that give the new value of the observables in terms of the old. The map $F_+$ takes $\{x_+(\sigma), x_+(\sigma)\}$ into $\{x_+(+ \vee \sigma), x_-(+ \vee \sigma)\}$ which is $\{a(1 + x_+), a x_-\}$ and the map $F_-$ takes $\{x_+, x_-\}$ into $\{a x_+, a(1 + x_-)\}$. In a more general case we have maps $F_\eta$ that take $x(\sigma) \mapsto x(\eta \vee \sigma)$, and a new operator $\mathcal{L}$

$$\mathcal{L} G(x) \stackrel{\text{def}}{=} \mathcal{T} G(x(\sigma)) = \sum_{\eta = \{+, -\}} G\left(F_\eta(x)\right) e^{-\beta x_\eta} \; . \quad (5)$$

Notice that all dependencies on $\sigma$ have disappeared — if we know the value of the state observables x the action of $\mathcal{L}$ on G can be computed.

The basic idea for the dynamics will be to choose the maps $F_\eta$ so that one of the state variables is the interaction energy. One can consider the two maps $F_+$ and $F_-$ as the inverse branches of a map f, that is, $f^{-1}(x) = \{F_+(x), F_-(x)\}$. Studying the thermodynamics of the interaction $\phi$ is equivalent to studying the long term behavior of the orbits of the map f, achieving the transformation of the spin system into a dynamical system.

Unlike the original transfer operator, the $\mathcal{L}$ operator — acting in the space of observables that depend only on the state variables — is of trace class (its trace is finite). The finite trace gives us a chance to relate the trace of $\mathcal{L}^n$ to the partition function of system of size $n$. We can do better. As most properties of interest (thermodynamics, fall-off of correlations) are determined directly from its spectrum, we can study instead the zeros of the Fredholm determinant $\det(1 - z\mathcal{L})$ by the technique of cycle expansions developed for dynamical systems [2]. A cycle expansion consists of finding a power series expansion for the determinant by writing $\det(1 - z\mathcal{L}) = \exp(\text{tr}\ln(1 - z\mathcal{L}))$. The logarithm is expanded into a power series and one is left with terms of the form $\text{tr}\mathcal{L}^n$ to evaluate. For evaluating the trace, the $\mathcal{L}$ operator is equivalent to

$$\mathcal{L}G(x) = \int_{\mathbf{R}} dy\, \delta(y - f(x)) e^{-\beta y} G(y) \qquad (6)$$

from which the trace can be computed:

$$\text{tr}\mathcal{L}^n = \sum_{x = f^{\circ n}(x)} \frac{e^{-\beta H(x)}}{|\det(1 - \partial_x f^{\circ n}(x))|} \qquad (7)$$

with the sum running over all the fixed points of $f^{\circ n}$ (all spin configurations of a given length). Here $f^{\circ n}$ is f composed with itself $n$ times, and $H(x)$ is the energy of the configuration associated with the point x. In practice the map f is never constructed and the energies are obtained directly from the spin configurations. The problem has now been reduced to a dynamical system and all the techniques applicable to the study of Fredholm determinants of dynamical systems can be used.

To compute the value of $\text{tr}\mathcal{L}^n$ we must compute the value of $\partial_x f^{\circ n}$; this involves a functional derivative. To any degree of accuracy a number x in the range of possible interaction energies can be represented by a finite string of spins $\epsilon$ as $x = \phi(+, \epsilon_0, \epsilon_1, \ldots, -, -, \ldots)$. By choosing the sequence $\epsilon$ to have a large sequence of spins $-$, the number x can be made as small as needed, so in particular we can represent a small variation $\delta = \phi(\eta)$. As $x_+(\epsilon) = \phi(+ \vee \epsilon)$, it can be computed by applying the map f to $\phi(\epsilon)$. From the definition of a derivative we have:

$$\partial_x f(x) = \lim_{m \to \infty} \frac{\phi(\epsilon \vee \eta^{(m)}) - \phi(\epsilon)}{\phi(\eta^{(m)})} \qquad (8)$$

where $\eta^{(m)}$ is a sequence of spin strings that make $\delta$ smaller and smaller. By substituting the definition of $\phi$ in terms of its pairwise interaction $J(n) = n^s a^{n^\gamma}$ and taking the limit for the sequences $\eta^{(m)} = \{+, -, -, \ldots, \eta_{m+1}, \eta_{m+2}, \ldots\}$ one obtains the limit is $a$ if $\gamma = 1$, 1 if $\gamma < 1$, and 0 if $\gamma > 1$. It does not depend on the positive value of s.

The manipulations have up to now assumed that the map f is smooth. If the dimension D of the embedding space is too small, f may not be smooth ($D = \infty$ cannot be ruled out). Determining under which conditions the embedding is smooth is a complicated question related to the problem of embedding an ergodic systems into classical systems [18]. There is no known complete solution to the problem, but in the case of spin systems with pairwise interaction it is possible to give a rule. If the interaction is of the form

$$\phi(\sigma) = \sum_{n \geq 1} \delta_{\sigma_0, \sigma_n} \sum_k p_k(n) a_k^{n^\gamma} \tag{9}$$

where $p_k$ are polynomials and $|a_k| < 1$, then for $\gamma \geq 1$ one can choose a dynamical system with dimension D equal to the sum of all degrees, that is $\sum \deg(p_k)$. The state observables to use in this case are $x_{s,k}(\sigma) = \sum \delta_{+, \sigma_n} n^s a_k^n$. An example is the interaction $J(n) = n^2(3/10)^n$. The action of the map $F_+$ for this interaction is illustrated figure 1. Plotted are the pairs $\{\phi(+ \vee \sigma), \phi(+ \vee + \vee \sigma)\}$. The plot does not represent a function, as there are points that have several images. A experimental point of view can be taken, and the plot considered as the chaotic attractor for an undetermined dynamical system. It should then be analyzed to determine the dimensionality of the dynamical system and how many dimensions are required to embedded it smoothly(what is D for the vector of state observables). Only the values of the interaction that actually occur as a value of a configuration are plotted, and in between the points one is free to connect them as required for the smoothness of the map $F_+$.

The added smoothness and trace class of the $\mathcal{L}$ operator translates into faster convergence towards the thermodynamic limit. If the reconstructed dynamics is analytic, then the convergence towards the thermodynamic limit will be faster than exponential [19, 20]. I will illustrate this with the polynomial-exponential interactions (9) with $\gamma = 1$; the convergence is faster than exponential if $\gamma > 1$. The convergence is illustrated in figure 2 for the interaction $n^2 e^{-\alpha n}$, which requires $D = 3$. Plotted in the graph to illustrate the transfer matrix convergence are the number of decimal digits that remain unchanged as the range of the interaction is increased. Also in the graph are the number of decimal digits that remain unchanged as the largest power of $\text{tr}\mathcal{L}^n$ considered. The plot is effectively a logarithmic plot and straight lines indicate exponentially fast convergence. The expansion for the $\mathcal{L}$ operator has curvature which indicates that the convergence is faster than exponential. By fitting, one can verify that the free energy is converging to its limiting value as $\exp(-n^{(4/3)})$. Cvitanović

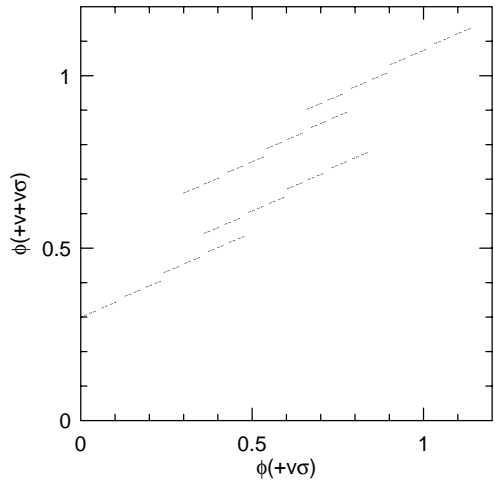

Figure 1: The spin adding map $F_+$ for the potential $J(n) = \sum n^2 a^{\alpha n}$. The action of the map takes the value of the interaction energy between $+$ and the semi-infinite configuration $\{\sigma_1, \sigma_2, \sigma_3, \ldots\}$ and returns the interaction energy between $+$ and the configuration $\{+, \sigma_1, \sigma_2, \sigma_3, \ldots\}$.

[19] has estimated that the Fredholm determinant of a map on a D dimensional space should converge as $\exp(-n^{(1+1/D)})$, which is confirmed by these numerical simulations.

The geometrization of spin systems strengthens the connection between statistical mechanics and dynamical systems. It also further establishes the value of the Fredholm determinant of the $\mathcal{L}$ operator as a practical computational tool with applications to chaotic dynamics, spin systems, and semiclassical mechanics. The example above emphasizes the high accuracy that can be obtained: by computing the shortest 14 periodic orbits of period 5 or less it is possible to obtain three digit accuracy for the free energy. For the same accuracy with a transfer matrix one has to consider a $256 \times 256$ matrix. This make the method of cycle expansions feasible for analytic calculations, and it also shows that a few configurations are dominating the behavior of the system.

There are many possibilities for developing the results presented. The cycle expansion can be factorized [3], the approach of Feigenbaum to the renormalization group can be used [21], and phase transitions can be explored as limiting case where the coefficients $a_k$ go to 1. The results also show that there is room for improvement in the convergence of Monte Carlo simulations. Rather than carrying out the averages in the spin systems, one can study averages of the dynamical system where convergence is faster.

I would like to acknowledge the hospitality of the Niels Bohr Institute and the Deutsche Bundesbahn where this work was carried out under the NATO/NSF post-doctoral fellowship RCD-9050092. It is also a pleasure to acknowledge discussions with Francis Alexander, Predrag Cvitanović, Robert Ecke, Brosl Hasslacher, and John Lowenstein.

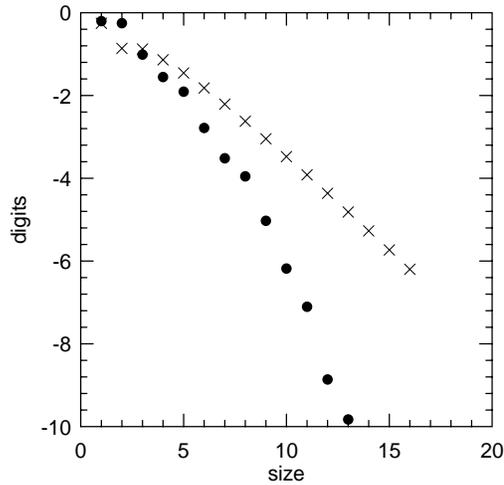

Figure 2: *Number of digits for the Fredholm method (•) and the transfer function method (×). The size refers to the largest cycle considered in the Fredholm expansions, and the truncation length in the case of the transfer matrix.*